\documentclass[10pt]{article}
\usepackage{graphicx}
\usepackage{amsmath}
\usepackage{amssymb}
\usepackage{caption2}
\begin{document}
\begin{center}
{\large {\bf \sc{  Strong decay  of the heavy tensor  mesons with  QCD sum rules }}} \\[2mm]
Zhi-Gang Wang \footnote{E-mail,zgwang@aliyun.com.  }    \\
 Department of Physics, North China Electric Power University,
Baoding 071003, P. R. China
\end{center}

\begin{abstract}
In the article, we  calculate the hadronic coupling constants $G_{D_2^*D\pi}$, $G_{D_{s2}^*DK}$, $G_{B_2^*B\pi}$, $G_{B_{s2}^*BK}$  with the three-point QCD sum rules, then study the two body strong decays $D_2^*(2460)\to D\pi$, $D_{s2}^*(2573)\to DK$, $B_2^*(5747)\to B\pi$, $B_{s2}^*(5840)\to BK$, and make predictions to be confronted with the experimental data in the future.
\end{abstract}



\section{Introduction}

The heavy-light mesons listed in the  Review of Particle Physics can be classified into the spin doublets in the heavy quark limit, now
the $\rm{1S}$ $(0^-,1^-)$ doublets $(B,B^*)$, $(D,D^*)$, $(B_s,B_s^*)$,  $(D_s,D_s^*)$ and
the $\rm{1P}$ $(1^+,2^+)$ doublets $(B_1(5721),B^*_2(5747))$,
 $(D_1(2420),D^*_2(2460))$, $(B_{s1}(5830),B^*_{s2}(5840))$, $(D_{s1}(2536),D^*_{s2}(2573))$ are complete \cite{PDG}.
   The doublet $(D_1(2420),D^*_2(2460))$ are  well-established experimentally, while the quantum numbers of the  $D^*_{s2}(2573)$ are not as well established,
  the width and decay modes are consistent with the $J^P=2^+$ assignment \cite{PDG}.
 In 2007, the D0 collaboration firstly observed the $B_1(5721)^0$ and $B_2(5747)^0$   \cite{D0-2007},
later the CDF   collaboration confirmed  them, and obtained the width $\Gamma(B_2^*)=\left(22.7^{+3.8}_{-3.2} {}^{+3.2}_{-10.2}\right) \,\rm{MeV}$  \cite{CDF-2008}.
Also in 2007, the CDF  collaboration observed the $B_{s1}(5830)$ and $B_{s2}^*(5840)$ \cite{CDF-Bs-2007}.
The D0 collaboration confirmed the  $B_{s2}^*(5840)$ \cite{D0-Bs-2007}.
In 2012, the LHCb collaboration updated the masses $M_{B_{s1}}=(5828.40\pm 0.04\pm 0.04\pm 0.41)\,\rm{ MeV}$ and $M_{B_{s2}^*}=(5839.99\pm 0.05\pm 0.11\pm 0.17)\,\rm{MeV}$, and measured the width $\Gamma(B_{s2}^*)=(1.56\pm0.13\pm0.47)\,\rm{MeV}$ \cite{LHCb-2012}. Recently, the CDF collaboration measured the masses and widths of the  $B_1(5721)$, $B^*_2(5747)$, $B_{s1}(5830)$, $B^*_{s2}(5840)$, and observed a new excited state $B(5970)$ \cite{CDF-2013}.

The  $\rm{1P}$ $(1^+,2^+)$ doublets have been drawn little attention compared to the $\rm{1S}$ $(0^-,1^-)$ and $\rm{1P}$ $(0^+,1^+)$ heavy-light mesons \cite{Swanson2006}.
We can study the masses, decay constants and strong decays of the $\rm{1P}$ $(1^+,2^+)$ doublets based on the QCD  sum rules to obtain fruitful information
about their internal structures and examine the heavy quark symmetry.  The P-wave, D-wave and radial excited heavy-light
mesons will be studied in details in the futures at the LHCb and KEK-B.
Experimentally, the strong decays of the $\rm{1P}$ $(1^+,2^+) $ doublets take place through relative D-wave,
  the corresponding widths are proportional to $|\vec{p}|^{2L+1}$, with the angular momentum $L=2$ transferred in the decays. In these decays, the momentum  $|\vec{p}|$ is small,  the decays are kinematically suppressed.
   The strong decays
 $B_1(5721)^0\to B^{*+}\pi^-$, $B_2(5747)^0 \to B^{*+}\pi^-,\,B^{+}\pi^-$ \cite{D0-2007,CDF-2008},
$B_{s1}(5830)^0 \to B^{*+}K^-$ \cite{CDF-Bs-2007,D0-Bs-2007,LHCb-2012}, $B_{s2}^*(5840)^0 \to B^{+}K^-$ \cite{CDF-Bs-2007,D0-Bs-2007,LHCb-2012}, $B_{s2}^*(5840)^0 \to B^{*+}K^-$ \cite{LHCb-2012},
$D_2^*(2460)^0\rightarrow D^{*+}\pi^-,\, D^+\pi^-$, $D_2^*(2460)^+\to D^0\pi^+$, $D_1(2420)^0\to D^{*+}\pi^-$, $D_1(2420)^+\to D^{*0}\pi^+$ \cite{PDG,BABAR-0808,BABAR-1009,LHCb1307}, $D_{s1}(2536)^+\to D^{*+}K^0,\,D^{*0}K^+$, $D_{s2}(2573)^+\to D^{0}K^+$ \cite{PDG}
 have been observed.

The  QCD sum rules (QCDSR) is a powerful nonperturbative theoretical tool in studying the
ground state hadrons, and has given many successful descriptions of the masses, decay constants, hadronic form-factors, hadronic coupling constants, etc  \cite{SVZ79,Reinders85,NarisonBook,ColangeloReview}.
The hadronic coupling constants in the
$D^*D \pi$,
$D^* D_s K$, $D_s^* D K$,
$B^*B \pi$,
$ B^{*}_{s}BK $,
$DD\rho$,
$D_{s}DK^*$, $ B_{s}BK^*$,
$D^{*}D \rho $, $D^{*}_{s}D K^{*}$, $B^{*}_{s}B K^{*}$,
$D^* D^* \rho$, $B^{*}B^{*}\rho $,
$B_{s0} B K$,  $B_{s1}B^{*}K$,
$ D^{*}_{s}D K_1 $, $ B^{*}_{s}BK_1 $,
$J/\psi D D$,
$J/\psi DD^*$,
$J/\psi D^*D^*$,
$ B_c^*B_c\Upsilon$, $ B_c^*B_c J/\psi$, $ B_cB_c\Upsilon$, $ B_cB_c J/\psi$
  vertices have been studied with the three-point QCDSR \cite{3PTQCDSR,3PTQCDSR-Rev},
while the hadronic coupling constants in the
  $D^* D \pi$, $D^*D_sK$, $D^*_sDK$, $B^* B \pi$,
    $DD\rho$, $DD_sK^*$, $D_sD_s\phi$, $BB\rho$,
    $D^{*}D\rho$, $D^*D_s K^*$, $D_s^*D_s \phi$, $B^*B\rho$,
   $D^*D^*\pi$, $D^*D_s^*K$,  $B^*B^*\pi$,
   $D^* D^* \rho$,
   $D_0D\pi$, $B_0B\pi$,  $D_0D_sK$, $D_{s0}DK$,   $B_{s0}BK$,
      $D_1D^* \pi$,   $B_1B^*\pi$, $D_{s1} D^* K$, $B_{s1} B^* K$, $B_1 B_0 \pi$,
    $B_2B_1\pi$, $B_2B^*\pi$,
    $B_1B^*\rho$, $B_1B\rho$, $B_2B^*\rho$, $B_2B_1\rho$ vertices have been studied with
the  light-cone QCDSR \cite{LCQCDSR}. The detailed knowledge  of the hadronic  coupling constants  is of great importance  in understanding  the effects of heavy quarkonium  absorptions in hadronic matter. Furthermore, the hadronic coupling constants
   play an important role in understanding
final-state interactions in the heavy quarkonium (or meson) decays and in other phenomenological analysis. Some hadronic coupling constants, such as  $G_{D_2^*D\pi}$, $G_{D_{s2}^*DK}$, $G_{B_2^*B\pi}$,
$G_{B_{s2}^*BK}$, can be directly extracted from the experimental data as the corresponding strong decays are kinematically allowed, we can confront the theoretical predications to the experimental data in the futures.

In Ref.\cite{HM-32-P}, K. Azizi et al study the masses and decay constants of the  tensor mesons $D_2^*(2460)$ and $D_{s2}^*(2573)$  with the QCDSR by only taking into account the perturbative terms and the mixed condensates  in the operator product expansion. In Ref.\cite{WangDi}, we calculate the contributions of the  vacuum condensates up to dimension-6 in the operator product expansion,  study the masses and decay constants of the heavy tensor mesons $D_2^*(2460)$, $D_{s2}^*(2573)$, $B_2^*(5747)$, $B_{s2}^*(5840)$ with the QCDSR.
The predicted masses of the $D_2^*(2460)$, $D_{s2}^*(2573)$, $B_2^*(5747)$, $B_{s2}^*(5840)$ are in excellent agreement with the experimental data, while the ratios of the decay constants $\frac{f_{D_{s2}^*}}{f_{D_{2}^*}}\approx\frac{f_{B_{s2}^*}}{f_{B_{2}^*}}\approx\frac{f_{D_{s}}}{f_{D}}\mid_{\rm exp}$, where the exp denotes the experimental value \cite{PDG}. In Ref.\cite{Azizi-1402},  K. Azizi et al calculate the hadronic   coupling constants
$g_{D_2^*D\pi}$ and $g_{D_{s2}^*DK}$  with the three-point QCDSR by choosing the tensor structure $p_{\mu}p_{\nu}$, then study the strong decays $D_{2}^{*}(2460)^{0}\rightarrow
D^+ \pi^- $ and $D_{s2}^{*}(2573)^{+}\rightarrow D^{+} K^{0}$, the decay widths are too small to account for  the experimental data, if the widths of the tensor mesons are saturated approximately by the two-body strong decays.
  In the article, we take the decay constants of the heavy tensor mesons as input parameters \cite{WangDi}, analyze  all the  tensor structures to study the vertices $D_2^*D\pi$, $D_{s2}^*DK$, $B_2^*B\pi$, $B_{s2}^*BK$  with the three-point QCDSR so as to choose the pertinent tensor structures (In this article, we choose the tensor structures $g_{\mu\nu}$ and $p^{\prime}_{\mu}p^{\prime}_{\nu}$, which differ from the tensor structure $p_{\mu}p_{\nu}$ chosen in Ref.\cite{Azizi-1402}.),  then obtain the corresponding hadronic coupling constants, and study the two-body strong decays $D_2^*(2460)\to D\pi$, $D_{s2}^*(2573)\to DK$, $B_2^*(5747)\to B\pi$, $B_{s2}^*(5840)\to BK$ and try to smear the large discrepancy between the theoretical calculations  and the experimental data \cite{Azizi-1402}.

The article is arranged as follows:  we derive the QCDSR for
the hadronic coupling constants in the  vertices $D_2^*D\pi$, $D_{s2}^*DK$, $B_2^*B\pi$, $B_{s2}^*BK$  in Sect.2;
in Sect.3, we present the numerical results and calculate the two body  strong decays; and Sect.4 is reserved for our
conclusions.

\section{QCD sum rules for  the hadronic coupling constants }
In the following, we write down  the three-point correlation functions
$\Pi_{\mu\nu}(p,p^\prime)$  in the QCDSR,
\begin{eqnarray}
\Pi_{\mu\nu}(p,p^\prime)&=&i^2\int d^4xd^4y e^{ip^\prime \cdot x}e^{i(p-p^\prime) \cdot (y-z)} \langle
0|T\left\{J_{\mathbb{D}}(x)J_{\mathbb{P}}(y)J_{\mu\nu}^{\dagger}(z)\right\}|0\rangle\mid_{z=0} \, , \\
J_{ \mathbb{D}}(x)&=& \overline{Q}(x)i\gamma_5q(x) \, , \nonumber\\
J_{ \mathbb{P}}(y)&=& \overline{q}(y)i\gamma_5q^{\prime}(y) \, , \nonumber\\
J_{\mu\nu}(z)&=&i\overline{Q}(z)\left( \gamma_\mu\stackrel{\leftrightarrow}{D}_\nu +\gamma_\nu\stackrel{\leftrightarrow}{D}_\mu-\frac{2}{3}\widetilde{g}_{\mu\nu} \stackrel{\leftrightarrow}{\!\not\!D}\right) q^{\prime}(z) \, , \\
\stackrel{\leftrightarrow}{D}_\mu&=&\left(\stackrel{\rightarrow}{\partial}_\mu-ig_sG_\mu\right)-\left(\stackrel{\leftarrow}{\partial}_\mu +ig_sG_\mu\right)\, ,\nonumber\\
\widetilde{g}_{\mu\nu}&=&g_{\mu\nu}-\frac{p_{\mu}p_{\nu}}{p^2} \, ,\nonumber
\end{eqnarray}
where $Q=c,b$ and $q,q^\prime=u,d,s$, the pseudoscalar currents $J_{ \mathbb{D}}(x)$ ($J_{ \mathbb{P}}(y)$) interpolate the heavy (light) pseudoscalar mesons $D$ and $B$ ($\pi$ and $K$), respectively,   the tensor currents $J_{\mu\nu}(z)$  interpolate the heavy tensor mesons $D_2^*(2460)$, $D_{s2}^*(2573)$, $B_2^*(5747)$ and $B_{s2}^*(5840)$, respectively.

We can insert  a complete set of intermediate hadronic states with
the same quantum numbers as the current operators $J_{\mu\nu}(0)$, $J_{ \mathbb{D}}(x)$ and $ J_{ \mathbb{P}}(y)$  into the
correlation functions $\Pi_{\mu\nu}(p,p^\prime)$  to obtain the hadronic representation
\cite{SVZ79,Reinders85}. After isolating the ground state
contributions from the heavy tensor mesons $\mathbb{T}$, heavy pseudoscalar mesons $\mathbb{D}$ and light pseudoscalar mesons $\mathbb{P}$, we get the following result,
\begin{eqnarray}
\Pi_{\mu\nu}(p,p^\prime)&=&\frac{f_{\mathbb{T}}M_{\mathbb{T}}^2f_{\mathbb{D}}M_{\mathbb{D}}^2f_{\mathbb{P}}M_{\mathbb{P}}^2 \,\,G_{\mathbb{TDP}}(q^2)}{(m_Q+m_q)(m_{q}+m_{q^\prime})\left(M_{\mathbb{T}}^2-p^2\right)\left(M_{\mathbb{D}}^2-p^{\prime2}\right)\left(M_{\mathbb{P}}^2-q^{2}\right)}
 \left\{ \frac{\lambda\left(M_\mathbb{T}^2,M_{\mathbb{D}}^2,q^2\right)}{12M_\mathbb{T}^2}g_{\mu\nu}\right.\nonumber\\
&&\left.+p^{\prime}_{\mu}p^{\prime}_{\nu}-\frac{M_\mathbb{T}^2+M_{\mathbb{D}}^2-q^2}{2M_\mathbb{T}^2}\left(p_{\mu}p^{\prime}_{\nu}+p^{\prime}_{\mu}p_{\nu} \right)+\left[ \frac{M_{\mathbb{D}}^2}{M_\mathbb{T}^2}+\frac{\lambda\left(M_\mathbb{T}^2,M_{\mathbb{D}}^2,q^2\right)}{6M_\mathbb{T}^4}\right]p_{\mu}p_{\nu}\right\}+\cdots\,  , \nonumber \\
&=&\Pi_1(p^2,p^{\prime2})g_{\mu\nu}+\Pi_2(p^2,p^{\prime2})p^{\prime}_{\mu}p^{\prime}_{\nu}+\Pi_3(p^2,p^{\prime2})\left(p_{\mu}p^{\prime}_{\nu}+p^{\prime}_{\mu}p_{\nu} \right)+\Pi_4(p^2,p^{\prime2})p_{\mu}p_{\nu}+\cdots \, , \nonumber\\
\end{eqnarray}
where $\lambda(a,b,c)=a^2+b^2+c^2-2ab-2bc-2ca$, the  decay constants $f_{\mathbb{T}}$, $f_{\mathbb{D}}$, $f_{\mathbb{P}}$ and the hadronic coupling constants $G_{\mathbb{TDP}}$ are defined by
\begin{eqnarray}
\langle 0|J_{\mu\nu}(0)|\mathbb{T}(p)\rangle&=&f_{\mathbb{T}}M^2_{\mathbb{T}}\,\varepsilon_{\mu\nu}  \, , \nonumber\\
\langle 0|J_{\mathbb{D}}(0)|\mathbb{D}(p^\prime)\rangle&=&\frac{f_{\mathbb{D}}M^2_{\mathbb{D}}}{m_Q+m_q}
\, , \nonumber\\
\langle 0|J_{\mathbb{P}}(0)|\mathbb{P}(q)\rangle&=&\frac{f_{\mathbb{P}}M^2_{\mathbb{P}}}{m_q+m_{q^\prime}} \, , \\
\langle {\mathbb{D}}(p^\prime){\mathbb{P}}(q)\mid \mathbb{T}(p)\rangle&=&G_{\mathbb{TDP}} \,\varepsilon_{\alpha\beta}(s,p)p^{\prime\alpha}q^\beta \, ,
\end{eqnarray}
the $\varepsilon_{\alpha\beta}$ are the polarization vectors of the tensor mesons  with the following properties,
\begin{eqnarray}
\sum_{s} \varepsilon^*_{\mu\nu}(s,p)\varepsilon_{\alpha\beta}(s,p)&=&\frac{\widetilde{g}_{\mu\alpha}\widetilde{g}_{\nu\beta}+\widetilde{g}_{\mu\beta}\widetilde{g}_{\nu\alpha}}{2}
-\frac{\widetilde{g}_{\mu\nu}\widetilde{g}_{\alpha\beta}}{3} \, .
\end{eqnarray}
In general, we expect that we can choose either component  $\Pi_{i}(p^2,p^{\prime2})$ (with $i=1,2,3,4$) of the correlations $\Pi_{\mu\nu}(p,p^\prime)$ to study the hadronic coupling constants $G_{\mathbb{TDP}}$. In calculations, we observe that the tensor structures $g_{\mu\nu}$ and $p_\mu^{\prime}p_\nu^{\prime}$ are the pertinent tensor structures. In Ref.\cite{Azizi-1402}, K. Azizi et al take the tensor currents $\hat{J}_{\mu\nu}(z)=i\overline{Q}(z)\left( \gamma_\mu\stackrel{\leftrightarrow}{D}_\nu +\gamma_\nu\stackrel{\leftrightarrow}{D}_\mu\right) q(z)$, which couple both to the heavy tensor mesons and heavy scalar mesons, some contaminations are introduced.

Now, we briefly outline  the operator product
expansion for the correlation functions $\Pi_{\mu\nu}(p,p^\prime)$  in perturbative
QCD.  We contract the quark fields in the correlation functions
$\Pi_{\mu\nu}(p,p^\prime)$ with Wick theorem firstly,
\begin{eqnarray}
\Pi_{\mu\nu}(p,p^\prime)&=&  \int   d^4xd^4y e^{ip^\prime \cdot x}e^{i(p-p^\prime) \cdot (y-z)}   {\rm Tr}\left\{i\gamma_{5}S^q_{ij}(x-y)i\gamma_{5}S^{q^\prime}_{jk}(y-z)\Gamma_{\mu\nu} S^Q_{ki}(z-x) \right\}\mid_{z=0}\, , \nonumber\\
\end{eqnarray}
where
\begin{eqnarray}
\Gamma_{\mu\nu}&=&i\left( \gamma_\mu\stackrel{\leftrightarrow}{\frac{\partial}{\partial z^\nu}} +\gamma_\nu\stackrel{\leftrightarrow}{\frac{\partial}{\partial z^\mu}}-\frac{2}{3}\widetilde{g}_{\mu\nu} \gamma^\tau\stackrel{\leftrightarrow}{\frac{\partial}{\partial z^\tau}}\right)\, ,
\end{eqnarray}
\begin{eqnarray}
S^{Q}_{ij}(x)&=&\frac{i}{(2\pi)^4}\int d^4k e^{-ik \cdot x} \left\{
\frac{\delta_{ij}}{\!\not\!{k}-m_Q}
-\frac{g_sG^n_{\alpha\beta}t^n_{ij}}{4}\frac{\sigma^{\alpha\beta}(\!\not\!{k}+m_Q)+(\!\not\!{k}+m_Q)
\sigma^{\alpha\beta}}{(k^2-m_Q^2)^2}\right.\nonumber\\
&&\left. +\frac{ig_s^2 GG\delta_{ij}}{12}\frac{m_Qk^2+m_Q^2\!\not\!{k}  }{(k^2-m_Q^2)^4}+\cdots\right\}\, ,
\end{eqnarray}
  $t^n=\frac{\lambda^n}{2}$, the $\lambda^n$ is the Gell-Mann matrix, the $i$, $j$, $k$ are color indexes \cite{Reinders85}. We usually choose the full light quark propagators in the coordinate space. In the present case, the quark condensates and mixed condensates have no contributions, so we take a simple replacement $Q\rightarrow q/q^\prime$ to obtain the full $q/q^\prime$ quark propagators. In the leading order approximation, the gluon field $G_\mu(z)$ in the covariant derivative has no contributions as $G_\mu(z)=\frac{1}{2}z^\lambda G_{\lambda\mu}(0)+\cdots=0$. Then we compute  the integrals to obtain the QCD spectral density through dispersion relation.

\begin{figure}
 \centering
 \includegraphics[totalheight=6cm,width=8cm]{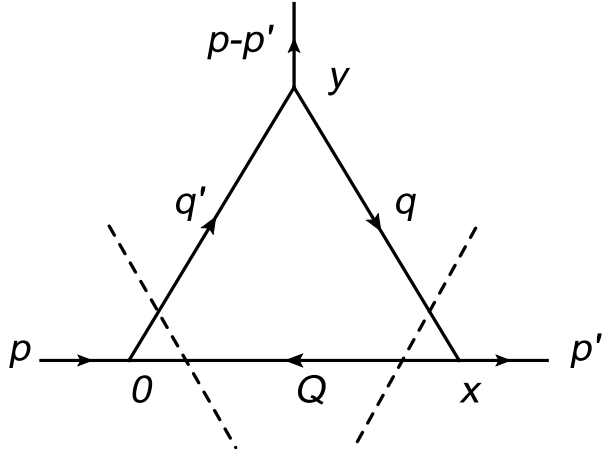}
    \caption{The leading-order   contributions, the dashed lines denote the Cutkosky's cuts.}
\end{figure}

\begin{figure}
 \centering
 \includegraphics[totalheight=8cm,width=15cm]{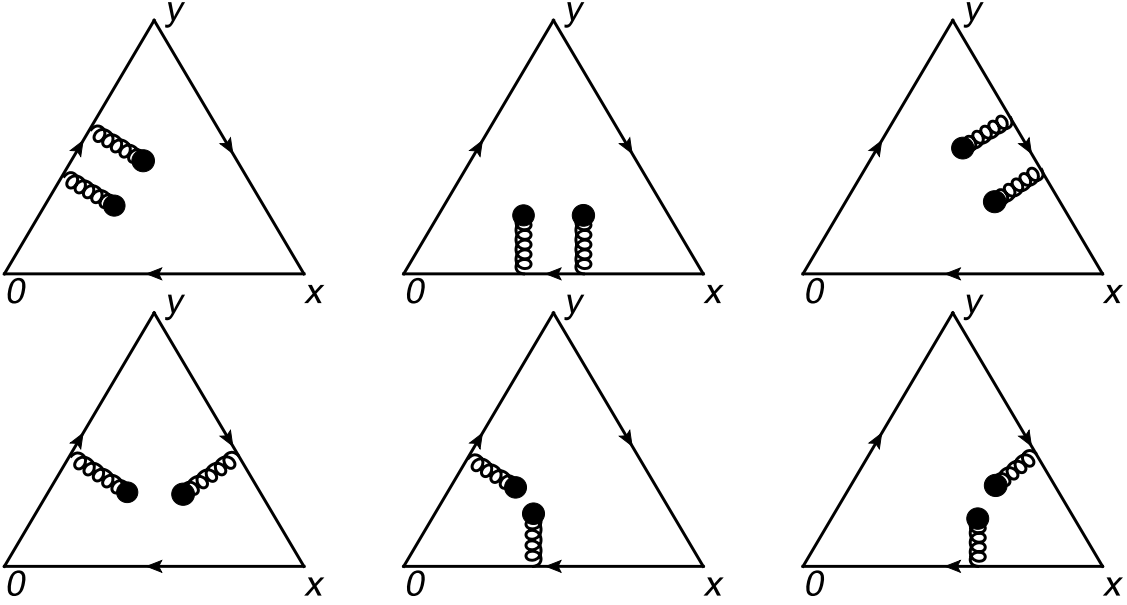}
    \caption{The gluon condensate   contributions. }
\end{figure}

The leading-order   contributions $\Pi_{\mu\nu}^{0}(p,p^{\prime})$ can be written as
\begin{eqnarray}
\Pi_{\mu\nu}^{0}(p,p^{\prime})&=&\frac{3i}{(2\pi)^4}\int d^4k \frac{ {\rm Tr}\left\{ \gamma_5\left[ \!\not\!{k}+m_q\right]\gamma_5 \left[ \!\not\!{k}+\!\not\!{p} -\!\not\!{p^{\prime}}+ m_{q^\prime}\right]\Gamma_{\mu\nu}\left[ \!\not\!{k}-\!\not\!{p^{\prime}} +m_Q\right]\right\}}{\left[k^2-m_q^2\right]\left[(k+p-p^{\prime})^2-m_{q^\prime}^2\right]\left[(k-p^{\prime})^2-m_Q^2\right]}\, ,\nonumber\\
&=&\int ds du \frac{\rho_{\mu\nu}}{(s-p^2)(u-p^{\prime2})} \, ,
\end{eqnarray}
where
\begin{eqnarray}
\Gamma_{\mu\nu}&=&\gamma_\mu(p-2k-2p^\prime)_\nu+\gamma_\nu(p-2k-2p^\prime)_\mu-\frac{2}{3}\widetilde{g}_{\mu\nu}\left(\!\not\!{p}+2\!\not\!{k}-2\!\not\!p^\prime\right)\, .
\end{eqnarray}
We put all the quark lines on mass-shell using the Cutkosky's rules, see Fig.1,
and  obtain the leading-order  spectral densities  $\rho_{\mu\nu}$,
\begin{eqnarray}
\rho_{\mu\nu}  &=&\frac{3}{(2\pi)^3} \int d^4k \delta\left[k^2-m_q^2\right]\delta\left[(k+p-p^{\prime})^2-m_{q^\prime}^2\right]\delta\left[(k-p^{\prime})^2-m_Q^2\right]\nonumber\\
&& {\rm Tr}\left\{ \gamma_5\left[ \!\not\!{k}+m_{q}\right]\gamma_5 \left[ \!\not\!{k}+\!\not\!{p} -\!\not\!{p^{\prime}}+ m_{q^\prime}\right]\Gamma_{\mu\nu}\left[ \!\not\!{k}-\!\not\!{p^{\prime}} +m_Q\right]\right\}   \, ,\\
&=&\frac{g_{\mu\nu}}{4\pi^2\sqrt{\lambda(s,u,q^2)}}\left\{ m_Q^3(m_{q^\prime}-m_q)-q^2m_Q(m_Q+m_{q^\prime})+m_Q(sm_q-um_{q^\prime})\right. \nonumber\\
&&\left.+ 6\left( u-s+q^2+2m_qm_Q-2m_{q^\prime}m_Q\right)d_2(0,0,m_Q)\right\} \nonumber\\
&&+\frac{3p^\prime_{\mu}p^\prime_{\nu}}{2\pi^2\sqrt{\lambda(s,u,q^2)}}\left\{ u+q^2-m_Q^2+2m_Qm_q\right. \nonumber\\
&&+(s-2u-2q^2+m_Q^2-4m_qm_Q+2m_{q^\prime}m_Q)b_1(0,0,m_Q) \nonumber\\
&&\left.+ \left( u-s+q^2+2m_qm_Q-2m_{q^\prime}m_Q\right)b_2(0,0,m_Q)\right\}+\cdots \, ,
\end{eqnarray}
where we have used the following formulae,
\begin{eqnarray}
\int d^4k\, \delta^3&=&\frac{\pi}{2\sqrt{\lambda(s,u,q^2)}} \, , \nonumber\\
\int d^4k\, \delta^3 \,k_\mu&=&\frac{\pi}{2\sqrt{\lambda(s,u,q^2)}}\left[a_1(m_A,m_B,m_Q)p_\mu+b_1(m_A,m_B,m_Q)p^\prime_\mu \right] \, , \nonumber\\
\int d^4k\, \delta^3 \, k_{\mu} k_\nu&=&\frac{\pi}{2\sqrt{\lambda(s,u,q^2)}}\left[a_2(m_A,m_B,m_Q)p_{\mu}p_\nu+b_2(m_A,m_B,m_Q)p^\prime_{\mu}p^\prime_{\nu} \right. \nonumber\\
&&\left. +c_2(m_A,m_B,m_Q)\left(p_{\mu}p^\prime_\nu+p^\prime_{\mu}p_\nu\right)+d_2(m_A,m_B,m_Q)g_{\mu\nu} \right]\, ,
\end{eqnarray}
\begin{eqnarray}
\delta^3&=& \delta\left[k^2-m_A^2 \right]\delta\left[(k+p-p^\prime)^2-m_B^2 \right]\delta\left[(k-p^\prime)^2-m_Q^2 \right] \, ,\nonumber\\
b_1(m_A,m_B,m_Q)&=& \frac{1}{\lambda(s,u,q^2)}\left[ m_Q^2(s-u+q^2)+u(u-s-2q^2)+q^2(q^2-s)\right.\nonumber\\
&&\left.-2sm_A^2+m_B^2(u+s-q^2)\right]\, , \nonumber\\
b_2(m_A,m_B,m_Q)&=&\frac{1}{\lambda(s,u,q^2)}\left[ (u-q^2-m_Q^2)^2+2m_B^2(u-q^2-m_Q^2)-4sm_A^2\right]\nonumber\\
&&+\frac{6s}{ \lambda^2(s,u,q^2)} \left\{ q^2\left[m_Q^4-(u+s-q^2)m_Q^2+su \right]+m_A^2m_B^2(q^2-u-s)\right.\nonumber\\
&&+m_A^2\left[ s(s-u-q^2)+m_Q^2(u-s-q^2)\right]\nonumber\\
&&\left.+m_B^2\left[ u(u-s-q^2)+m_Q^2(s-u-q^2)\right]\right\}\, , \nonumber\\
d_2(m_A,m_B,m_Q)&=& \frac{1}{ 2\lambda(s,u,q^2) } \left\{ q^2\left[m_Q^4-(u+s-q^2)m_Q^2+su \right]+m_A^2m_B^2(q^2-u-s)\right.\nonumber\\
&&+m_A^2\left[ s(s-u-q^2)+m_Q^2(u-s-q^2)\right]\nonumber\\
&&\left.+m_B^2\left[ u(u-s-q^2)+m_Q^2(s-u-q^2)\right]\right\}\, ,
\end{eqnarray}
here we have neglected the terms $m_A^4$ and $m_B^4$ as they are irreverent in present calculations.
The gluon condensate contributions shown by the Feynman diagrams in Fig.2 are calculated accordingly.

We take quark-hadron duality below the continuum thresholds $s_0$ and $u_0$ respectively, and perform the double Borel transform  with respect to the variables
$P^2=-p^2$ and $P^{\prime2}=-p^{\prime2}$ to obtain the QCDSR,

\begin{eqnarray}
\Pi_1(M_1^2,M_2^2)&=&\frac{f_{\mathbb{T}}M_{\mathbb{T}}^2f_{\mathbb{D}}M_{\mathbb{D}}^2f_{\mathbb{P}}M_{\mathbb{P}}^2 \,\,G_{\mathbb{TDP}}(q^2)}{(m_Q+m_q)(m_{q}+m_{q^\prime})\left(M_{\mathbb{P}}^2-q^{2}\right)}
  \frac{\lambda\left(M_\mathbb{T}^2,M_{\mathbb{D}}^2,q^2\right)}{12M_\mathbb{T}^2}\exp\left( -\frac{M_\mathbb{T}^2}{M_1^2}-\frac{M_{\mathbb{D}}^2}{M_2^2}\right)\nonumber\\
  &=&\int ds du \exp\left(-\frac{s}{M_1^2}-\frac{u}{M_2^2}\right)\left\{\frac{1}{4\pi^2\sqrt{\lambda(s,u,q^2)}}\left[ m_Q^3(m_{q^\prime}-m_q)-q^2m_Q(m_Q+m_{q^\prime})\right.\right. \nonumber\\
&&\left.+m_Q(sm_q-um_{q^\prime})+ 6\left( u-s+q^2+2m_qm_Q-2m_{q^\prime}m_Q\right)d_2(0,0,m_Q)\right] \nonumber\\
&&+\frac{1}{\sqrt{\lambda(s,u,q^2)}}\langle\frac{\alpha_sGG}{\pi}\rangle\left[\frac{1}{9s}-\frac{s-u-3q^2}{12}\frac{\partial^2}{\partial m_A^2\partial m_B^2}d_2(m_A,m_B,m_Q)\right.\nonumber\\
&&-\frac{s-3u-q^2}{12}\frac{\partial^2}{\partial m_A^2\partial m_Q^2}d_2(m_A,0,m_Q)+\frac{s+u+q^2}{12}\frac{\partial^2}{\partial m_B^2\partial m_Q^2}d_2(0,m_B,m_Q)\nonumber\\
&&\left.\left.-\frac{1}{3}\frac{\partial}{\partial m_A^2}d_2(m_A,0,m_Q)-\frac{1}{2}\frac{\partial}{\partial m_B^2}d_2(0,m_B,m_Q)-\frac{1}{2}\frac{\partial}{\partial m_Q^2}d_2(0,0,m_Q)\right]\right\}\, ,
\end{eqnarray}

 \begin{eqnarray}
\Pi_2(M_1^2,M_2^2)&=&\frac{f_{\mathbb{T}}M_{\mathbb{T}}^2f_{\mathbb{D}}M_{\mathbb{D}}^2f_{\mathbb{P}}M_{\mathbb{P}}^2 \,\,G_{\mathbb{TDP}}(q^2)}{(m_Q+m_q)(m_{q}+m_{q^\prime})\left(M_{\mathbb{P}}^2-q^{2}\right)}
   \exp\left( -\frac{M_\mathbb{T}^2}{M_1^2}-\frac{M_{\mathbb{D}}^2}{M_2^2}\right)\nonumber\\
  &=&\int ds du \exp\left(-\frac{s}{M_1^2}-\frac{u}{M_2^2}\right)\left\{\frac{3}{2\pi^2\sqrt{\lambda(s,u,q^2)}}\left[ u+q^2-m_Q^2+2m_Qm_q\right.\right. \nonumber\\
&&+\left(s-2u-2q^2+m_Q^2-4m_qm_Q+2m_{q^\prime}m_Q\right)b_1(0,0,m_Q)\nonumber\\
&&\left.+ \left( u-s+q^2+2m_qm_Q-2m_{q^\prime}m_Q\right)b_2(0,0,m_Q)\right] \nonumber\\
&&+\frac{1}{\sqrt{\lambda(s,u,q^2)}}\langle\frac{\alpha_sGG}{\pi}\rangle\left[ -\frac{s-u-3q^2}{12}\frac{\partial^2}{\partial m_A^2\partial m_B^2}b_2(m_A,m_B,m_Q)\right.\nonumber\\
&&-\frac{s-3u-q^2}{12}\frac{\partial^2}{\partial m_A^2\partial m_Q^2}b_2(m_A,0,m_Q)+\frac{s+u+q^2}{12}\frac{\partial^2}{\partial m_B^2\partial m_Q^2}b_2(0,m_B,m_Q)\nonumber\\
&&-\frac{1}{3}\frac{\partial}{\partial m_A^2}b_2(m_A,0,m_Q)-\frac{1}{2}\frac{\partial}{\partial m_B^2}b_2(0,m_B,m_Q)-\frac{1}{2}\frac{\partial}{\partial m_Q^2}b_2(0,0,m_Q) \nonumber\\
&&\left.\left.+\frac{5}{6}\frac{\partial}{\partial m_A^2}b_1(m_A,0,m_Q)+\frac{11}{12}\frac{\partial}{\partial m_B^2}b_1(0,m_B,m_Q)+\frac{11}{12}\frac{\partial}{\partial m_Q^2}b_1(0,0,m_Q)\right]\right\} \, , \nonumber\\
\end{eqnarray}
where
\begin{eqnarray}
\int dsdu&=&\int_{m_Q^2}^{s_0} ds \int_{m_Q^2}^{u_0}du \mid_{-1\leq\frac{\left(u-q^2-m_Q^2\right)\left(s+u-q^2\right)-2s\left(u-m_Q^2\right)}{|u-q^2-m_Q^2|\sqrt{\lambda(u,s,q^2)}}\leq 1} \,\, ,
\end{eqnarray}
\begin{eqnarray}
\frac{\partial^2}{\partial m_i^2 \partial m_j^2 } f(m_A,m_B,m_Q)\doteq \frac{\partial^2}{\partial m_i^2 \partial m_j^2 } f(m_A,m_B,m_Q)\mid_{m_A=0;m_B=0}\, , \nonumber\\
\frac{\partial}{\partial m_i^2  } f(m_A,m_B,m_Q)\doteq \frac{\partial}{\partial m_i^2  } f(m_A,m_B,m_Q)\mid_{m_A=0;m_B=0}\, ,
\end{eqnarray}
and $f(m_A,m_B,m_Q)=b_1(m_A,m_B,m_Q)$, $b_2(m_A,m_B,m_Q)$, $d_2(m_A,m_B,m_Q)$, $\cdots$, $m_i^2,m_j^2=m_A^2$, $m_B^2$, $m_Q^2$.

\section{Numerical results and discussions}
The hadronic input parameters are taken as
$M_{D^{*}_2(2460)^\pm}=(2464.3\pm1.6)\,\rm{MeV}$,
$M_{D^{*}_2(2460)^0}=(2461.8\pm0.7)\,\rm{MeV}$,
$M_{D^{*}_{s2}(2573)}=(2571.9\pm0.8)\,\rm{MeV}$,
$M_{B^{*}_2(5747)^0}=(5743\pm5)\,\rm{MeV}$,
$M_{B^{*}_{s2}(5840)^0}=(5839.96\pm0.20)\,\rm{MeV}$,
$M_{D^\pm}=(1869.5 \pm 0.4)\,\rm{MeV}$,
$M_{D^0}=(1864.91 \pm 0.17)\,\rm{MeV}$,
$M_{B^\pm}=(5279.25\pm0.26)\,\rm{MeV}$,
$M_{B^0}=(5279.55\pm0.26)\,\rm{MeV}$,
$M_{K^\pm}=(493.677\pm0.013)\,\rm{MeV}$,
$M_{K^0}=(497.614\pm0.022)\,\rm{MeV}$,
$M_{\pi^\pm}=(139.57018\pm0.00035)\,\rm{MeV}$,
$M_{\pi^0}=(134.9766\pm0.0006)\,\rm{MeV}$,
 $f_\pi=130\,\rm{MeV}$, $f_K=156\,\rm{MeV}$ from the Particle Data Group
\cite{PDG}.
The threshold parameters  are taken as $s^0_{D^*_2}=(8.5\pm0.5)\,\rm{GeV}^2$, $s^0_{D^*_{s2}}=(9.5\pm0.5)\,\rm{GeV}^2$, $s^0_{B^*_2}=(39\pm 1)\,\rm{GeV}^2$, $s^0_{B^*_{s2}}=(41\pm 1)\,\rm{GeV}^2$, $u^0_{D}=(6.2\pm0.5)\,\rm{GeV}^2$, $u^0_{B}=(33.5\pm1.0)\,\rm{GeV}^2$ from the QCDSR \cite{WangDi,WangJHEP}. Then the energy gaps $\sqrt{s_0/u_0}-M_{\rm ground \, state}=(0.4-0.6)\,\rm{GeV}$, the contributions of the ground states are fully included.

 The value of the gluon condensate $\langle \frac{\alpha_s
GG}{\pi}\rangle $ is taken as the standard  value $\langle \frac{\alpha_s GG}{\pi}\rangle=0.012 \,\rm{GeV}^4 $ \cite{ColangeloReview}. The masses the $u$ and $d$ quarks are obtained through  the Gell-Mann-Oakes-Renner relation
$f_{\pi}^2m_{\pi}^2=2(m_u+m_d)\langle\bar{q}q\rangle$, i.e. $m_u=m_d=6\,\rm{MeV}$ at the energy scale $\mu=1\,\rm{GeV}$.

In the article, we take the $\overline{MS}$ masses  $m_{c}(m_c)=(1.275\pm0.025)\,\rm{GeV}$, $m_{b}(m_b)=(4.18\pm0.03)\,\rm{GeV}$ and $m_s(\mu=2\,\rm{GeV})=(0.095\pm0.005)\,\rm{GeV}$
 from the Particle Data Group \cite{PDG}, and take into account
the energy-scale dependence of  the $\overline{MS}$ masses from the renormalization group equation,
\begin{eqnarray}
m_s(\mu)&=&m_s({\rm 2GeV} )\left[\frac{\alpha_{s}(\mu)}{\alpha_{s}({\rm 2GeV})}\right]^{\frac{4}{9}} \, ,\nonumber\\
m_c(\mu)&=&m_c(m_c)\left[\frac{\alpha_{s}(\mu)}{\alpha_{s}(m_c)}\right]^{\frac{12}{25}} \, ,\nonumber\\
m_b(\mu)&=&m_b(m_b)\left[\frac{\alpha_{s}(\mu)}{\alpha_{s}(m_b)}\right]^{\frac{12}{23}} \, ,\nonumber\\
\alpha_s(\mu)&=&\frac{1}{b_0t}\left[1-\frac{b_1}{b_0^2}\frac{\log t}{t} +\frac{b_1^2(\log^2{t}-\log{t}-1)+b_0b_2}{b_0^4t^2}\right]\, ,
\end{eqnarray}
  where $t=\log \frac{\mu^2}{\Lambda^2}$, $b_0=\frac{33-2n_f}{12\pi}$, $b_1=\frac{153-19n_f}{24\pi^2}$, $b_2=\frac{2857-\frac{5033}{9}n_f+\frac{325}{27}n_f^2}{128\pi^3}$,  $\Lambda=213\,\rm{MeV}$, $296\,\rm{MeV}$  and  $339\,\rm{MeV}$ for the flavors  $n_f=5$, $4$ and $3$, respectively  \cite{PDG}.
  In Ref.\cite{WangDi},  we study the masses and decay constants of the heavy tensor mesons using the QCDSR, and obtain the values
  $M_{D_2^*}=(2.46\pm0.09)\,\rm{GeV}$,
$M_{D_{s2}^*}=(2.58\pm0.09)\,\rm{GeV}$,
$M_{B_2^*}=(5.73\pm0.06)\,\rm{GeV}$,
$M_{B_{s2}^*}=(5.84\pm0.06)\,\rm{GeV}$,
$f_{D_2^*}=(0.182\pm0.020)\,\rm{GeV}$,
$f_{D_{s2}^*}=(0.222\pm0.021)\,\rm{GeV}$,
$f_{B_2^*}=(0.110\pm0.011)\,\rm{GeV}$,
$f_{B_{s2}^*}=(0.134\pm0.011)\,\rm{GeV}$.
The predicted masses $M_{D_2^*}$,
$M_{D_{s2}^*}$,
$M_{B_2^*}$ and
$M_{B_{s2}^*}$ are in excellent agreement with the experimental data.

  In calculations, we take $n_f=4$ and  $\mu=  1(3)\,\rm{GeV}$ for the charmed (bottom) tensor mesons \cite{WangDi}, and evolve all the scale dependent quantities to the energy scales  $\mu=  1 \,\rm{GeV}$ and $\mu=  3 \,\rm{GeV}$ respectively through the renormalization group equation. The same energy scales and truncations in the operator product expansion lead to the values $M_D=1.87\,\rm{GeV}$, $M_B=5.28\,\rm{GeV}$,  $f_D=156\,\rm{MeV}$ and $f_{B}=168\,\rm{MeV}$. If we take into account the perturbative corrections, the experimental values $f_{D}=205\,\rm{MeV}$ and $f_{B}=190\,\rm{MeV}$ can be reproduced \cite{PDG,WangJHEP,Rosner}.  In this article, we take the values of the decay constants of the heavy-light mesons as $f_{D_2^*}=0.182\,\rm{GeV}$,
$f_{D_{s2}^*}=0.222\,\rm{GeV}$,
$f_{B_2^*}=0.110\,\rm{GeV}$,
$f_{B_{s2}^*}=0.134\,\rm{GeV}$, $f_D=0.156\,\rm{GeV}$ and $f_{B}=0.168\,\rm{GeV}$, and neglect the uncertainties so as to avoid doubling counting as the uncertainties originate mainly from the threshold parameters and heavy quark masses.

From the QCDSR in Eqs.(16-17), we can see that there are no contributions come from the quark condensates and mixed condensates, and no terms of the orders ${\mathcal{O}}\left(\frac{1}{M_1^2}\right)$,  ${\mathcal{O}}\left(\frac{1}{M_2^2}\right)$,  ${\mathcal{O}}\left(\frac{1}{M_1^4}\right)$,  ${\mathcal{O}}\left(\frac{1}{M_2^4}\right)$, $\cdots$, which are needed to stabilize the QCDSR so as to warrant a platform.
 In this article, we take the local limit $M_1^2=M_2^2\rightarrow \infty$, and obtain the local QCDSR.
 The ground states, higher resonances and continuum states have the same weight $\exp\left(-M_{\mathbb{T}}^2/M_1^2 -M_{\mathbb{D}}^2/M_2^2\right)=1$, we use the
 threshold parameters (or the cut-off) $s_0$ and $u_0$ to avoid the contaminations of the  higher resonances and continuum states, while the threshold parameters $s_0$ and $u_0$ are determined by the conventional QCDSR \cite{WangDi}. At the QCD side, there are not terms of the orders ${\mathcal{O}}\left(\frac{1}{M_1^2}\right)$,  ${\mathcal{O}}\left(\frac{1}{M_2^2}\right)$,  ${\mathcal{O}}\left(\frac{1}{M_1^4}\right)$,  ${\mathcal{O}}\left(\frac{1}{M_2^4}\right)$, which vanish in the limit $M_1^2=M_2^2\rightarrow \infty$, so the threshold parameters $s_0$ and $u_0$ survive in the local QCDSR.

Now we obtain the hadronic coupling constants $G_{\mathbb{TDP}}(q^2=-Q^2)$ at the large space-like regions, for example, $Q^2\geq 3\,\rm{GeV}^2$, then fit the hadronic coupling constants $G_{\mathbb{TDP}}(Q^2)$ into the functions $A_i+B_{i}Q^2$, where $i=\rm{C,\,U,\,L}$, the C, U, and L denote the central values, upper bound and lower bound, respectively, the numerical  values are shown the Table 1.
If the heavy quark symmetry and chiral symmetry work well, the physical values of the hadronic coupling constants  should have the relations,
\begin{eqnarray}
\frac{G_{D_{s2}^*DK}(Q^2=-M_{K}^2)}{G_{D_2^*D\pi}(Q^2=-M_{\pi}^2)}\approx \frac{G_{B_{s2}^*BK}(Q^2=-M_{K}^2)}{G_{B_2^*B\pi}(Q^2=-M_{\pi}^2)}&\approx& 1\, .
\end{eqnarray}
From Table 1, we can see that the ratio,
\begin{eqnarray}
\frac{G_{D_{s2}^*DK}(Q^2=-M_{K}^2)}{G_{D_2^*D\pi}(Q^2=-M_{\pi}^2)}\approx \frac{G_{B_{s2}^*BK}(Q^2=-M_{K}^2)}{G_{B_2^*B\pi}(Q^2=-M_{\pi}^2)}&\approx& \frac{3}{4}\, ,
\end{eqnarray}
which is smaller than the expectation 1.
In calculations, we have used the $s$-quark mass $m_s=95\,\rm{MeV}$ at the energy scale $\mu=2\,\rm{GeV}$, if we take larger value (the value of the $m_s$ varies  in a rather large range \cite{ColangeloReview}), say $m_s=130\,\rm{MeV}$, the relations in Eq.(21) can be satisfied. So in this article, we prefer the values $G_{D_2^*D\pi}(Q^2=-M_{\pi}^2)$ and $G_{B_2^*B\pi}(Q^2=-M_{\pi}^2)$ from the QCDSR as they suffer from much less uncertainties induced by the light quark masses, and   take the approximation
$G_{D_{s2}^*DK}(Q^2=-M_{K}^2)=G_{D_2^*D\pi}(Q^2=-M_{\pi}^2)$ and $G_{B_{s2}^*BK}(Q^2=-M_{K}^2)=G_{B_2^*B\pi}(Q^2=-M_{\pi}^2)$ according to the heavy quark symmetry and chiral symmetry.

The perturbative QCD spectral densities associate with the tensor structure $g_{\mu\nu}$ have dimension (of mass) 2, while the perturbative QCD spectral densities associate with the tensor structure $p^\prime_{\mu}p^\prime_{\nu}$ have dimension 0, it is more reliable to take the perturbative QCD spectral densities associate with the tensor structure $g_{\mu\nu}$ as they can embody the energy dependence efficiently.   The values of the hadronic coupling constants come  from the QCDSR associate with the tensor $g_{\mu\nu}$ are much larger than that of the tensor $p^\prime_{\mu}p^\prime_{\nu}$.  In this article, we prefer the values $ G_{D_2^*D\pi}(Q^2=-M_{\pi}^2)=16.5^{+3.3}_{-3.5}\,\rm{GeV}^{-1}$, $
G_{B_2^*B\pi}(Q^2=-M_{\pi}^2)=39.3^{+4.9}_{-5.2}\,\rm{GeV}^{-1}$ associate with the tensor $g_{\mu\nu}$, as they can also lead to much larger decay widths and favor accounting  for  the experimental data.

 \begin{table}
\begin{center}
\begin{tabular}{|c|c|c|c|c|c|c|}\hline\hline
   $g_{\mu\nu}$                               & $D_2^*D\pi$           & $D_{s2}^*DK$           & $B_2^*B\pi$            & $B_{s2}^*BK$           \\ \hline
  $Q^2  $                                     & $3.0-5.0$             & $3.0-5.0$              & $3.5-5.5$              & $3.5-5.5$           \\ \hline
  $A_{\rm C}$                                 & $16.42481$            & $11.92224$             & $39.18672$             & $25.67374$           \\
  $B_{\rm C}$                                 & $-1.86478$            & $-1.23275$             & $-3.98713$             & $-2.3704$           \\ \hline
  $A_{\rm U}$                                 & $19.74325$            & $14.18738$             & $44.15991$             & $28.71525$           \\
  $B_{\rm U}$                                 & $-1.99324$            & $-1.30484$             & $-4.00222$             & $-2.34827$      \\ \hline
  $A_{\rm L}$                                 & $12.96084$            & $9.55968$              & $33.97408$             & $22.48229$           \\
  $B_{\rm L}$                                 & $-1.67737$            & $-1.12313$             & $-3.89453$             & $-2.34741$      \\ \hline
  $G_{\mathbb{TDP}}(Q^2=-M_{\mathbb{P}}^2)$   & $16.5^{+3.3}_{-3.5}$  & $12.2^{+2.3}_{-2.4}$   & $39.3^{+4.9}_{-5.2}$   & $26.3^{+3.0}_{-3.2}$        \\ \hline
     \hline
  $p^\prime_{\mu}p^\prime_{\nu}$              & $D_2^*D\pi$           & $D_{s2}^*DK$           & $B_2^*B\pi$            & $B_{s2}^*BK$            \\ \hline
  $Q^2 $                                      & $3.0-5.0$             & $3.0-5.0$              & $3.5-5.5$              & $3.5-5.5$           \\ \hline
  $A_{\rm C}$                                 & $12.31645$            & $9.69653$              & $17.07687$             & $12.66033$           \\
  $B_{\rm C}$                                 & $-1.3785$             & $-0.99737$             & $-1.64969$             & $-1.12767$           \\ \hline
  $A_{\rm U}$                                 & $14.90752$            & $11.57224$             & $19.45758$             & $14.31228$           \\
  $B_{\rm U}$                                 & $-1.47863$            & $-1.0608$              & $-1.64827$             & $-1.11951$      \\ \hline
  $A_{\rm L}$                                 & $9.58102$             & $7.71456$              & $14.55604$             & $10.90844$           \\
  $B_{\rm L}$                                 & $-1.2291$             & $-0.90211$             & $-1.60863$             & $-1.10864$      \\ \hline
  $G_{\mathbb{TDP}}(Q^2=-M_{\mathbb{P}}^2)$   & $12.3^{+2.6}_{-2.7}$  & $9.9^{+1.9}_{-2.0}$    & $17.1^{+2.4}_{-2.5}$   & $12.9^{+1.7}_{-1.7}$        \\ \hline
     \hline
\end{tabular}
\end{center}
\caption{ The parameters of the hadronic coupling constants $G_{\mathbb{TDP}}(Q^2)$, where the $g_{\mu\nu}$ and $p^\prime_{\mu}p^\prime_{\nu}$ denote the   tensor structures of the QCDSR, the units of the $G_{\mathbb{TDP}}(Q^2)$, $A_i$, $B_i$ and $Q^2$ are $\rm{GeV}^{-1}$, $\rm{GeV}^{-1}$, $\rm{GeV}^{-2}$ and $\rm{GeV}^{2}$, respectively.    }
\end{table}

We can take the hadronic coupling constants $G_{\mathbb{TDP}}(Q^2=-M^2_{\mathbb{P}})$ as basic input parameters and study  the following strong decays,
\begin{eqnarray}
D_2^*(2460)  &\to& D^+\pi^-, \,D^0\pi^0 \, , \nonumber\\
D_{s2}^*(2573) &\to& D^0K^+, \,D^+K^0 \, , \nonumber\\
B_2^*(5747) &\to& B^+\pi^- , \, B^0\pi^0\, , \nonumber\\
B_{s2}^*(5840) &\to& B^+K^-,\,B^0\bar{K}^0\, ,
\end{eqnarray}
which take place through relative D-wave. The decay widths can be written as
\begin{eqnarray}
\Gamma&=&C_p\frac{G^2_{\mathbb{TDP}}|\vec{p}|^5}{60\pi M_{\mathbb{T}}^2} \, ,
\end{eqnarray}
where
\begin{eqnarray}
|\vec{p}|&=&\frac{\sqrt{\lambda\left(M_{\mathbb{T}}^2,M_{\mathbb{D}}^2, M_{\mathbb{P}}^2 \right)}}{2M_{\mathbb{T}}}\, ,\nonumber
\end{eqnarray}
 $C_p=1$ (or $\frac{1}{2}$) for the final states $\pi^\pm$, $K$ (or $\pi^0$).
The numerical results are
\begin{eqnarray}
\Gamma(D_2^*(2460)  \to D^+\pi^-)&=&7.91^{+3.49}_{-3.00}\,\rm{MeV}\, , \nonumber\\
\Gamma(D_2^*(2460)  \to D^0\pi^0)&=&4.14^{+1.82}_{-1.57}\,\rm{MeV}\, ,  \nonumber\\
\Gamma(D_{s2}^*(2573) \to D^0K^+)&=&3.35^{+1.48}_{-1.27}\,\rm{MeV}\, , \nonumber\\
\Gamma(D_{s2}^*(2573) \to D^+K^0)&=&3.04^{+1.34}_{-1.15}\,\rm{MeV}\, , \nonumber \\
\Gamma(B_2^*(5747) \to B^+\pi^-)&=&3.42^{+0.90}_{-0.85}\,\rm{MeV}\, , \nonumber\\
\Gamma(B_2^*(5747) \to B^0\pi^0)&=&1.73^{+0.46}_{-0.43}\,\rm{MeV}\, , \nonumber \\
\Gamma(B_{s2}^*(5840) \to B^+K^-)&=&0.25^{+0.06}_{-0.06}\,\rm{MeV}\, ,\nonumber\\
\Gamma(B_{s2}^*(5840) \to B^0\bar{K}^0)&=&0.21^{+0.06}_{-0.05}\,\rm{MeV}\, .
\end{eqnarray}
From the experimental data of the BaBar collaboration,
\begin{eqnarray}
\frac{\Gamma(D_2^*(2460) \rightarrow D^+\pi^-)}{\Gamma(D_2^*(2460) \rightarrow D^+\pi^-)+\Gamma(D_2^*(2460) \rightarrow
D^{*+}\pi^-)}&=&0.62\pm0.03\pm0.02\, \cite{BABAR-0808}\, , \nonumber
\end{eqnarray}
\begin{eqnarray}
\frac{\Gamma(D_2^*(2460) \rightarrow D^+\pi^-)}{\Gamma(D_2^*(2460) \rightarrow
D^{*+}\pi^-)}&=&1.47 \pm 0.03 \pm 0.16\, \cite{BABAR-1009}\, ,
\end{eqnarray}
we can obtain the average,
\begin{eqnarray}
\frac{\Gamma(D_2^*(2460) \rightarrow D^+\pi^-)}{\Gamma(D_2^*(2460) \rightarrow
D^{*+}\pi^-)}&=&1.55\, ,
\end{eqnarray}
which is consistent with the PDG's average $1.54\pm0.15$ \cite{PDG}. We assume
\begin{eqnarray}
\frac{\Gamma(D_2^*(2460) \rightarrow D^0\pi^0)}{\Gamma(D_2^*(2460) \rightarrow
D^{*0}\pi^0)}&=&\frac{\Gamma(D_2^*(2460) \rightarrow D^+\pi^-)}{\Gamma(D_2^*(2460) \rightarrow
D^{*+}\pi^-)}=1.55\, ,
\end{eqnarray}
and saturate the total decay width $\Gamma(D_2^*(2460))$ with the two-body strong decays $D_2^*(2460)  \to D^+\pi^-$, $D^{*+}\pi^-$, $D^0\pi^0$, $D^{*0}\pi^0$,
then obtain the theoretical value,
\begin{eqnarray}
\Gamma(D_2^*(2460)^0 )&=&(12-29)\,\rm{MeV} \, ,
\end{eqnarray}
 which is much smaller than the experimental value,
\begin{eqnarray}
\Gamma(D_2^*(2460)^0 )&=&(49.0\pm 1.3)\,\rm{MeV} \,\,\, \,\, \,\,\, \,\, \,\,\, \,\, \,\,\, \,{\rm PDG's \,\,  average } \,\,\cite{PDG}\, , \nonumber\\
&=&(43.2 \pm 1.2 \pm 3.0)\, {\rm MeV}\,\,\,\,{\rm from \,\, the \,\, final \,\, state}\,\, D^{*+}\pi^-\, \, \cite{LHCb1307} \, , \nonumber\\
&=&(45.6 \pm 0.4 \pm 1.1)\, {\rm MeV}\,\,\,\,{\rm from \,\, the \,\, final \,\, state}\,\, D^{+}\pi^-\, \, \cite{LHCb1307} \, .
\end{eqnarray}

The strong decays $D_{s2}^*(2573) \to  D^{*0}K^+,  \,D^{*+}K^0$ are greatly suppressed in the phase-space, while the strong decays
$D_{s2}^*(2573) \to   D_s^+\pi^0,\,D_s^{*+}\pi^0$ violate the isospin conservation and are also greatly suppressed. We saturate the total decay width $\Gamma(D_{s2}^*(2573) )$ with the two-body strong decays $D_{s2}^*(2573) \to D^0K^+$, $D^+K^0$, and obtain the  theoretical value,
\begin{eqnarray}
\Gamma(D_{s2}^*(2573) )&=&(4-9)\,\rm{MeV}\, ,
\end{eqnarray}
 which is smaller than the experimental value,
\begin{eqnarray}
\Gamma(D_{s2}^*(2573) )&=&(17 \pm4)\,\rm{MeV}\, \cite{PDG}\, .
\end{eqnarray}

At the bottom sector, we assume  $\Gamma(B_2^*(5747)\to B^*\pi) = \Gamma(B_2^*(5747)\to  B\pi)$ according to the experimental value \cite{PDG} \begin{eqnarray}
\frac{\Gamma(B_2^*(5747)\to B^*\pi)}{\Gamma(B_2^*(5747)\to  B \pi)} &=& 1.10\pm 0.42\pm 0.31\,  ,
\end{eqnarray}
 and neglect the  kinematically suppressed decays
 $B_{s2}^*(5840)\to B^{*+}K^-,\,B^{*0}\bar{K}^0$ and isospin violated decays  $B_{s2}^*(5840)\to B_s^0\pi^0,\,B_s^{*0}\pi^0$,
and saturate the total decay widths $\Gamma(B_2^*(5747))$ and $\Gamma(B_{s2}^*(5840))$ with the two-body strong decays   $ B_2^*(5747)\to B^+\pi^- , \, B^{*+}\pi^-, \,B^0\pi^0, \,B^{*0}\pi^0$ and $B_{s2}^*(5840)\to B^+K^-,\,B^0\bar{K}^0$, respectively. Then
we obtain the theoretical values,
\begin{eqnarray}
\Gamma(B_2^*(5747)^0)&=&(8-13)\,\rm{MeV}\, ,\nonumber\\
\Gamma(B_{s2}^*(5840)) &=&(0.4-0.6)\,\rm{MeV}\, ,
\end{eqnarray}
  which are smaller than the experimental values,
\begin{eqnarray}
\Gamma(B_2^*(5747)^0)&=&(26 \pm 3 \pm 3)\,\rm{MeV} \, \cite{CDF-2013} \, ,\nonumber\\
\Gamma(B_{s2}^*(5840))&=&(2.0 \pm 0.4 \pm 0.2)\, \rm{MeV}\, \cite{CDF-2013} \, .
\end{eqnarray}

The perturbative $\mathcal{O}({\alpha_s})$ corrections increase the correlation function (or the product $f_Bf_{B^*}G_{B^*B\pi}$) about $50\%$ in the light-cone QCD sum rules for the hadronic coupling constant $G_{B^*B\pi}$ \cite{Kho-99}. In the present case, we can assume the perturbative $\mathcal{O}({\alpha_s})$ corrections also increase the correlation functions (or the products $f_{\mathbb{T}}f_{\mathbb{D}}G_{\mathbb{TDP}}$) about $50\%$. The perturbative $\mathcal{O}({\alpha_s})$ corrections to the decay constants
 $f_{\mathbb{T}}$ are negative \cite{WangDi}, the net perturbative $\mathcal{O}({\alpha_s})$ corrections to the $f_{\mathbb{D}}G_{\mathbb{TDP}}$ are larger  than $50\%$. If half of those perturbative $\mathcal{O}({\alpha_s})$ corrections are compensated by the perturbative $\mathcal{O}({\alpha_s})$ corrections to the decay constants $f_{\mathbb{D}}$,  the hadronic coupling constants $G_{\mathbb{TDP}}$ are increased by about $30\%$, then taking into account
the perturbative $\mathcal{O}({\alpha_s})$ corrections lead to the following replacements,
 \begin{eqnarray}
 G_{\mathbb{TDP}} &\to& 1.3G_{\mathbb{TDP}}   \, , \nonumber\\
  \Gamma(D_2^*(2460)^0 ) &\to& (20-49)\,\rm{MeV}  \, , \nonumber\\
  \Gamma(D_{s2}^*(2573) )&\to&(7-15)\,\rm{MeV}\, ,\nonumber\\
  \Gamma(B_2^*(5747)^0)&\to&(14-22)\,\rm{MeV}\, ,\nonumber\\
\Gamma(B_{s2}^*(5840) ) &\to&(0.7-1.0)\,\rm{MeV}\, .
 \end{eqnarray}
   Then the theoretical values $\Gamma(D_2^*(2460)^0 )$, $ \Gamma(D_{s2}^*(2573))$ and $ \Gamma(B_2^*(5747)^0)$ are   compatible with the experimental data, while the theoretical value $ \Gamma(B_{s2}^*(5840))$ is still smaller than the experimental value.

\section{Conclusion}
In the article, we choose the pertinent tensor structures to calculate the hadronic coupling constants $G_{D_2^*D\pi}$, $G_{D_{s2}^*DK}$, $G_{B_2^*B\pi}$, $G_{B_{s2}^*BK}$  with the three-point QCDSR,  then study the two body strong decays $D_2^*(2460)\to D\pi$, $D_{s2}^*(2573)\to DK$, $B_2^*(5747)\to B\pi$, $B_{s2}^*(5840)\to BK$, the predicted total widths are compatible with the experimental data, while the predicted partial widths can be confronted with the experimental data  from the BESIII, LHCb,  CDF, D0 and KEK-B  collaborations  in the futures. We can also take the hadronic coupling constants  as basic input parameter in  many phenomenological analysis.

\section*{Acknowledgements}
This  work is supported by National Natural Science Foundation,
Grant Numbers 11375063, the Fundamental Research Funds for the
Central Universities,  and Natural Science Foundation of Hebei province, Grant Number A2014502017.

\end{document}